\documentclass[twocolumn]{aastex6}

\usepackage{bm}
\usepackage{amsmath}
\usepackage{xcolor}
\usepackage{graphicx}
\usepackage{hyperref}
\usepackage[all]{hypcap} 

\begin{document}
	\title{Large ion-neutral drift velocities and plasma heating in partially ionized coronal rain blobs}
	\shorttitle{Ion-neutral drifts in coronal rain}
	\shortauthors{Martínez-Gómez et al.}
	
	\author{David Martínez-Gómez\altaffilmark{1,4}, Ramón Oliver\altaffilmark{2,3}, Elena Khomenko\altaffilmark{1,4}, Manuel Collados\altaffilmark{1,4}}
	\altaffiltext{1}{Instituto de Astrofísica de Canarias, 38205 La Laguna, Tenerife, Spain}
	\altaffiltext{2}{Departament de Física, Universitat de les Illes Balears, 07122, Palma de Mallorca, Spain}
	\altaffiltext{3}{Institut d'Aplicacions Computacionals de Codi Comunitari (IAC3), Universitat de les Illes Balears, 07122, Palma de Mallorca, Spain}
	\altaffiltext{4}{Departamento de Astrofísica, Universidad de La Laguna, 38205 La Laguna, Tenerife, Spain}
	\email{dmartinez@iac.es}

	\begin{abstract}
		In this paper we present a numerical study of the dynamics of partially ionized coronal rain blobs. We use a two-fluid model to perform a high-resolution 2D simulation that takes into account the collisional interaction between the charged and neutral particles contained in the plasma. We follow the evolution of a cold plasma condensation as it falls through an isothermal vertically stratified atmosphere that represents the much hotter and lighter solar corona. We study the consequences of the different degrees of collisional coupling that are present in the system. On the one hand, we find that at the dense core of the blob there is a very strong coupling and the charged and neutral components of the plasma behave as a single fluid, with negligible drift velocities (of a few $\rm{cm \ s^{-1}}$). On the other hand, at the edges of the blob the coupling is much weaker and larger drift velocities (of the order of $1 \ \rm{km \ s^{-1}}$) appear. In addition, frictional heating causes large increases of temperature at the transition layers between the blob and the corona. For the first time we show that such large drift velocities and temperature enhancements can develop as a consequence of ion-neutral decoupling associated to coronal rain dynamics. This can lead to enhanced emission coming from the plasma at the coronal rain-corona boundary, which possesses transition region temperature.
	\end{abstract}
	\keywords{plasmas -- magnetohydrodynamics (MHD) -- Sun: corona -- Sun: filaments, prominences, coronal rain}

\section{Introduction}
    Plasma in the solar atmosphere may be found in a state of partial ionization. This means that it may contain a non-negligible amount of neutral particles which are not directly affected by the presence of magnetic fields. Neutral species have been detected, for instance, in solar prominences \citep[see, e.g.,][]{Hirayama1985SoPh..100..415H}, in chromospheric spicules \citep{Pereira2016ApJ...824...65P} or in coronal rain blobs \citep{Antolin2012ApJ...745..152A,Schad2018ApJ...865...31S}.
	
    From the theoretical point of view it is known that the dynamics of the ionized and the neutral species decouple at small enough spatial and temporal scales. This has been shown, for instance, by the works of \citet{Hillier2019PhPl...26h2902H} and \citet{Popescu2021A&A...646A..93P,Popescu2021A&A...650A.181P}, who used multi-fluid models to numerically study the development of the Kelvin-Helmholtz instability (KHI) and the Rayleigh-Taylor instability (RTI), respectively, in prominences. It has also been shown that the different behavior of the charged and neutral particles leads to the damping of waves and plasma heating due to the collisions between the various species \citep[see, e.g.,][]{Leake2005A&A...442.1091L,Zaqarashvili2011A&A...529A..82Z,Soler2013ApJ...767..171S,MartinezGomez2018ApJ...856...16M,Popescu2019A&A...630A..79P}
	 
    However, the observational evidence regarding the different dynamics of charged and neutral species in the solar atmosphere is still scarce, mainly due to the requirements about temporal and spatial resolution. \citet{Khomenko2016ApJ...823..132K} studied the Doppler shifts of the spectral lines He \textsc{I} 10830 $\rm{\AA}$  and Ca \textsc{II} 8542 $\rm{\AA}$ in observations of a solar prominence and deduced the existence of large ion-neutral drift velocities (of the order of several hundred $\rm{m \ s^{-1}}$) in small-scale and short-lived transients. \citet{Anan2017A&A...601A.103A} analyzed a larger set of spectral lines and also found differences in the Doppler velocities but they concluded that they were not caused by decoupling effects but were the result of observing motions originated at different regions of the prominence. \citet{Stellmacher2017SoPh..292...83S} performed observations of a quiescent prominence and systematically found an excess of $\sim 10\%$ of the ion velocity over the neutral velocity. These results were later confirmed by \citet{Wiehr2019ApJ...873..125W,Wiehr2021ApJ...920...47W}. Recent studies by \citet{Gonzalez-Manrique2022} and \citet{Zapior2022ApJ...934...16Z} also point to the existence of large drift velocities (of up $1.7 \ \rm{km \ s^{-1}}$) in small regions of solar prominences. Regarding coronal rain observations, \citet{Ahn2014SoPh..289.4117A} compared the Doppler velocities computed from the spectral lines $\rm{H\alpha}$ and Ca \textsc{II} 8542 $\rm{\AA}$ and found that they ``matched well'', although their results show that they are not completely identical.
	 
    Coronal rain is a very appropriate phenomenon to study the interaction between charged and neutrals species \citep{Oliver2016ApJ...818..128O}. It involves a large range of ionization degree of the plasma, from the fully ionized case of the solar corona to the partially or weakly ionized condition of the falling blobs, which causes a large variation in the collisional coupling between the various species. However, most of the works on coronal rain dynamics have only considered the fully ionized case. This applies both to one-dimensional \citep[e.g.,][]{Muller2005A&A...436.1067M, Mendoza-Briceno2005ApJ...624.1080M, Antolin2010ApJ...716..154A, Mikic2013ApJ...773...94M, Froment2018ApJ...855...52F} and multi-dimensional numerical simulations \citep[e.g.][]{Mok2016ApJ...817...15M, Xia2017A&A...603A..42X, Kohutova2020A&A...639A..20K,Li2022ApJ...926..216L}. 
    
    In the present work, we are interested in investigating the effects of that ion-neutral collisional interaction during the fall of cold condensations of plasma towards the solar surface. Therefore, we use a two-fluid model \citep{Khomenko2014PhPl...21i2901K,Popescu2019A&A...627A..25P} to perform high-resolution 2D simulations of the evolution of partially ionized coronal rain blobs. We focus on analyzing the existence of charged-neutral drift velocities and the plasma heating caused by the collisional friction. Then, we discuss observational consequences of these processes.

\section{Methodology}
\subsection{Model}
    In this work, we assume that the plasma is composed of hydrogen only. Therefore, it contains three different kinds of particles: electrons, protons, and neutrals, which are denoted by the subscripts `e', `p', and `n', respectively. We also assume that the plasma fulfills the charge quasi-neutrality condition, so the number density of electrons and protons is the same, that is, $n_{\rm{e}} = n_{\rm{p}}$. Furthermore, a strong coupling between electrons and protons is considered, which allows to treat them together as a single charged fluid, denoted by the subscript `c'. Then, the temporal evolution of this plasma is described by a two-fluid 2D model in which charged and neutral particles interact by means of elastic collisions. 
    
    The details of the full set of equations included in the two-fluid model can be found in \citet{Popescu2019A&A...627A..25P} and \citet{Martinez2021A&A...650A.123M}. Here we only show the most relevant terms for the present investigation, which are the momentum transfer and heating due to collisions, given by
	\begin{equation} \label{eq:Rcoll}
		\bm{R}_{\rm{cn}} = \alpha \rho_{\rm{c}} \rho_{\rm{n}} \left(\bm{V}_{\rm{n}} - \bm{V}_{\rm{c}} \right)
	\end{equation}
	and
	\begin{equation} \label{eq:qv}
		Q_{V} = \alpha \rho_{\rm{c}}\rho_{\rm{n}}\frac{V_{\rm{D}}^{2}}{2},
	\end{equation}
	respectively. In the previous expressions, the variables $\rho_{\rm{c}}$, $\rho_{\rm{n}}$, $\bm{V}_{\rm{c}}$, and $\bm{V}_{\rm{n}}$ represent the mass density of charges, the mass density of neutrals, the velocity of charges, and the velocity of neutrals, respectively. The parameter $\alpha$ is known as the collisional coupling coefficient and is given by
	\begin{equation} \label{eq:alpha_def}
		\alpha = \frac{\rho_{\rm{e}} \nu_{\rm{en}} + \rho_{\rm{p}} \nu_{\rm{pn}}}{\rho_{\rm{n}} \rho_{\rm{c}}},
	\end{equation}
	where $\nu_{\rm{en}}$ and $\nu_{\rm{pn}}$ are the electron-neutral and proton-neutral collision frequencies. The frequency of collisions between two different species `s' and `t' (when one of them is neutral) can be computed from the following expression \citep{Draine1986MNRAS.220..133D}:
	\begin{equation} \label{eq:nu_st}
		\nu_{\rm{st}} = \frac{n_{\rm{t}} m_{\rm{t}}}{m_{\rm{s}}+m_{\rm{t}}} \sqrt{\frac{8 k_{\rm{B}}}{\pi}\left(\frac{T_{\rm{s}}}{m_{\rm{s}}}+\frac{T_{\rm{t}}}{m_{\rm{t}}}\right)} \sigma_{\rm{st}}.
	\end{equation}
	Here, $n_{\rm{s}}$,  $m_{\rm{s}}$, and $T_{\rm{s}}$ represent the number density, mass, and temperature of the species `s', respectively. The parameter $k_{B}$ is the Boltzmann constant and $\sigma_{\rm{st}}$ is the collisional cross-section between the two species. We use the following values, taken from \citet{Leake2013PhPl...20f1202L}: $\sigma_{\rm{pn}} = 1.16 \times 10^{-18} \ \rm{m^{2}}$ and $\sigma_{\rm{en}} = 10^{-19} \ \rm{m^{2}}$. We note that $n_{\rm{c}} = n_{\rm{e}} + n_{\rm{p}} = 2 n_{\rm{p}}$, and the relation between the number and mass densities are $\rho_{\rm{c}} = \rho_{\rm{p}} + \rho_{\rm{e}} = n_{\rm{p}} m_{\rm{p}} + n_{\rm{e}} m_{\rm{e}} \approx n_{\rm{p}} m_{\rm{p}} = n_{\rm{c}}/2 m_{\rm{p}}$ and $\rho_{\rm{n}} = n_{\rm{n}} m_{\rm{n}} \approx n_{\rm{n}} m_{\rm{p}}$ (since for the case of hydrogen plasma $m_{\rm{n}} \approx m_{\rm{p}}$).
	Finally, the variable $V_{\rm{D}}$ is the modulus of the vector $\bm{V}_{\rm{D}}$, which represents the drift velocity between the two fluids and is given by
	\begin{equation} \label{eq:vd}
		\bm{V}_{\rm{D}} = (V_{\rm{c,x}}-V_{\rm{n,x}},V_{\rm{c,z}}-V_{\rm{n,z}})=\left(V_{\rm{D,x}}, V_{\rm{D,z}}\right).
	\end{equation}

\subsection{Numerical setup}
	We numerically solve the two-fluid equations using the \textsc{Mancha-2F} code \citep{Popescu2019A&A...627A..25P}. In this code, the physical variables, $\bm{f}$, are treated as the sum of equilibrium values ($\bm{f}_{\rm{eq}}$, which remain constant in time) and perturbation values ($\bm{f}_{1}$, which evolve with time and can be non-linear). 
	
	To simulate the fall of a partially ionized coronal rain blob we use as equilibrium a 2D isothermal vertically stratified atmosphere, with a temperature given by $T_{0} = 2 \times 10^{6} \ \rm{K}$, a uniform vertical magnetic field given by $B_{0} = 5 \ \rm{G}$, and a plasma ionization degree given by $\chi_{\rm{c},eq} = \rho_{\rm{c},eq}/\rho_{T,\rm{eq}} = 0.9$ (where $\rho_{T,\rm{eq}} = \rho_{\rm{c},eq} + \rho_{\rm{n},eq}$). We note that in reality the equilibrium atmosphere representing the solar corona should be fully ionized (that is, $\chi_{\rm{c},eq} = 1$). However, due to numerical reasons (the fluid equations cannot be solved when $\rho_{\rm{s}}(x,z) = 0$, since these variables appear in the denominators of several terms), we need to include a small fraction of neutrals in this environment. Thus, for our simulation we set the relative abundance of neutrals in the equilibrium state as $\chi_{\rm{n},eq}=\rho_{\rm{n},eq}/\rho_{T,\rm{eq}} = 1 - \chi_{\rm{c},eq} = 0.1$.
	
	As shown by \citet{Oliver2016ApJ...818..128O}, the condition of magneto-hydrostatic equilibrium leads to the following solutions from the two-fluid model:
	\begin{gather} \label{eq:equil}
	    \rho_{\rm{c},eq}(z) = \rho_{\rm{c},0} e^{-z/H_{\rm{c}}}, \ \rho_{\rm{n},eq}(z) = \rho_{\rm{n},0} e^{-z/H_{\rm{n}}},\\
	    P_{\rm{c},eq}(z) = P_{\rm{c},0} e^{-z/H_{\rm{c}}}, \ P_{\rm{n},eq}(z) = P_{\rm{n},0} e^{-z/H_{\rm{n}}},
	\end{gather}
	where $P_{\rm{c},eq}$ and $P_{\rm{n},eq}$ are the equilibrium pressures of the charged and neutral fluids, respectively, and the following relations, corresponding to the values of density and pressure at the base of the corona ($z = 0$), are satisfied:
	\begin{equation} \label{eq:p_rho_rel}
	    P_{\rm{c},0} = \frac{2 \rho_{\rm{c},0}k_{\rm{B}}T_{0}}{m_{\rm{p}}}, \ P_{\rm{n},0} = \frac{ \rho_{\rm{n},0}k_{\rm{B}}T_{0}}{m_{\rm{p}}}.
	\end{equation}
	The parameters $H_{\rm{c}}$ and $H_{\rm{n}}$ are the vertical scale heights of the charged and neutral fluids, respectively, and are computed as
	\begin{equation} \label{eq:hc_hn}
	    H_{\rm{c}} = \frac{2k_{\rm{B}} T_{0}}{m_{\rm{p}} g} \ \text{and} \ H_{\rm{n}} = \frac{k_{\rm{B}}T_{0}}{m_{\rm{p}}g}.
	\end{equation}
	For the simulation analyzed in the present work we use a value of $g = 273.98 \ \rm{m \ s^{-2}}$ for the solar surface gravity, which produces vertical scale heights of $H_{\rm{c}} \approx 120 \ \rm{Mm}$ and $H_{\rm{n}} \approx 60 \ \rm{Mm}$. In addition, the densities at the coronal base are given by $\rho_{\rm{c},0} = \chi_{\rm{c},eq} \rho_{T,0}$ and $\rho_{\rm{n},0} = (1 - \chi_{\rm{c},eq}) \rho_{T,0}$, where $\rho_{T,0} = 5 \times 10^{-12} \ \rm{kg \ m^{-3}}$ is the total density.
	
    We represent the coronal rain blob as a density perturbation for each species `s' given by
	\begin{equation} \label{eq:blob_profile}
	    \rho_{\rm{s},1}(x,z) = \chi_{\rm{s}} \rho_{b0} \exp \left[-\frac{\left(x - x_{0}\right)^{2} + \left(z - z_{0}\right)^{2}}{\Delta^{2}}\right],
	\end{equation}
	where $\chi_{\rm{n}} = (1 - \chi_{\rm{c}})$, $\rho_{b0}$ is the total density at the center of the blob, $x_{0}$ and $z_{0}$ are its initial coordinates, and $\Delta$ determines its width. For the present work we consider the following values of these parameters: $\rho_{b0} = 10^{-9} \ \rm{kg \ m^{-3}}$, $\chi_{\rm{c}} = 0.1$ (the blob is weakly ionized), $x_{0} = 0$, $z_{0} = 50 \ \rm{Mm}$, and $\Delta = 0.5 \ \rm{Mm}$. The chosen value for $\rho_{b0}$ produces a blob to corona density ratio of $\sim 300$ at the height $z_{0}$. Taking into account this density ratio and the equilibrium in pressures, the temperature of the blob is $T_{b0} \approx 7000 \ \rm{K}$, which corresponds to a typically chromospheric temperature.
	
\begin{figure}
	    \centering
	    \includegraphics[width=\hsize]{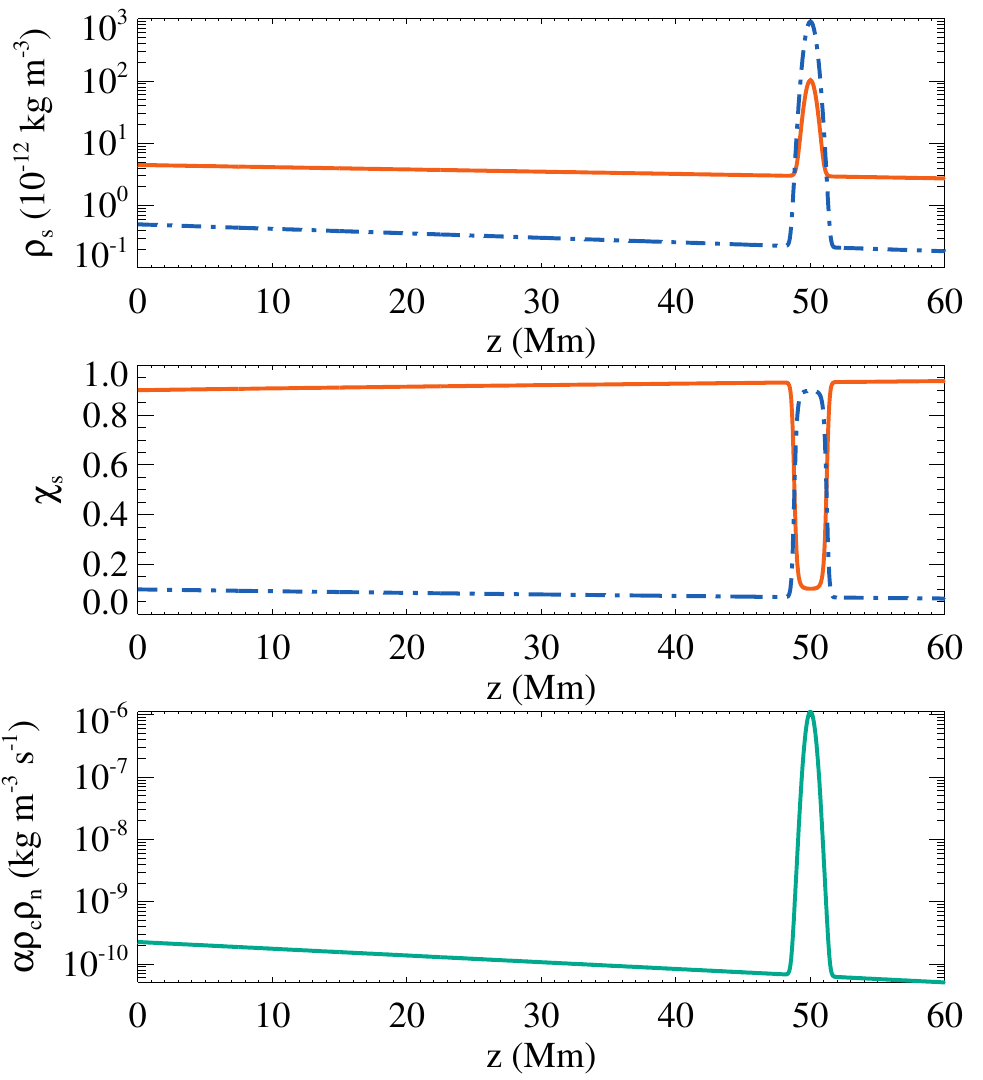}
	    \caption{Sketch of the initial time of the simulation. Top and middle panels show the densities, $\rho_{\rm{s}}$, and the relative abundances, $\chi_{\rm{s}}$, of the charged and neutral fluids, represented by orange solid lines and blue dashed-dotted lines, respectively. Bottom panel shows the coupling degree between the two fluids, given by the term $\alpha \rho_{\rm{c}} \rho_{\rm{n}}$.}
	    \label{fig:init1d}
	\end{figure}

	Therefore, at the initial time of the simulation, the density profiles are given by the sum of Eqs. (\ref{eq:equil}) and (\ref{eq:blob_profile}), as shown in the top panel of Fig. \ref{fig:init1d} (which corresponds to vertical cuts of densities at the position $x = 0$). In addition, the middle panel of Fig. \ref{fig:init1d} shows the relative abundances of the two fluids, $\chi_{\rm{c}}$ and $\chi_{\rm{n}}$, and the bottom panel shows the collisional coupling between the two fluids (represented by the factor $\alpha \rho_{\rm{c}} \rho_{\rm{n}}$) as a function of height.
	
	Finally, we consider a physical domain that extends from $-5 \ \rm{Mm}$ to $5 \ \rm{Mm}$ in the horizontal direction and from $-20 \ \rm{Mm}$ to $80 \ \rm{Mm}$ in the vertical direction. We represent it with a numerical mesh of $10^{3} \times 10^{4}$ points, which corresponds to a resolution of $10 \ \rm{km}$. We apply periodic boundary conditions in the horizontal direction and Perfectly Matched Layers \citep[PML;][]{Berenger1994JCoPh.114..185B,Parchevsky2009ApJ...694..573P} in the vertical direction.
	
\section{Results}
\begin{figure*}
    	\centering
    	\includegraphics[width=\hsize]{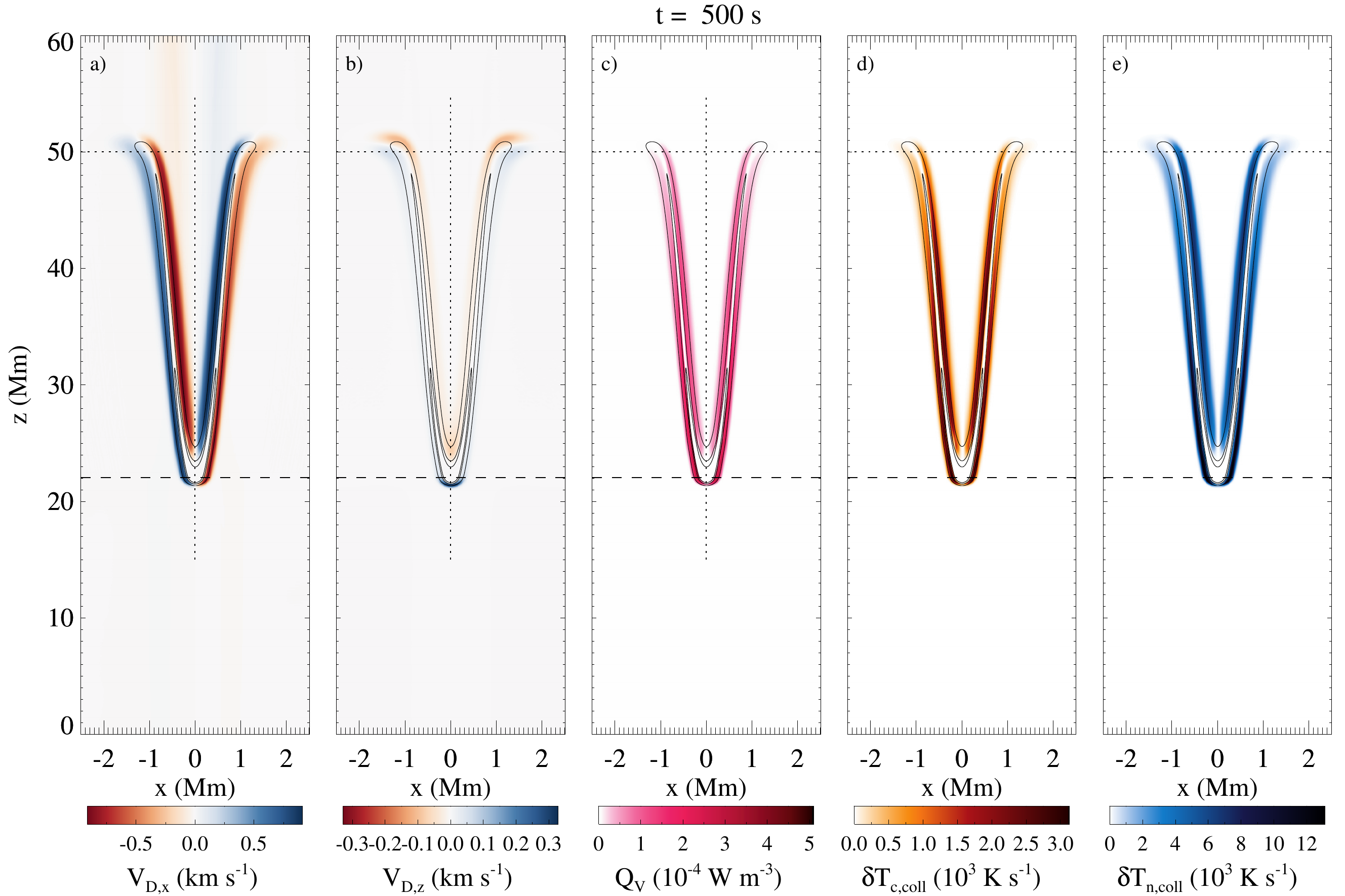}
    	\caption{Snapshot of the simulation of a falling coronal rain blob. \textit{a)} and \textit{b):} color maps of the $x$-component and $z$-component of the drift velocity, $V_{\rm{D},x}$ and $V_{\rm{D},z}$, respectively; \textit{c)}: color map of the frictional heating, $Q_{V}$; \textit{d)} and \textit{e):} color maps of the temperature variations of the charged and the neutral fluid, respectively, caused by the frictional heating represented in \textit{c)}. Black solid lines represent density contours of the neutral fluid. A vertical dotted line has been drawn at $x = 0$ and two horizontal dashed and dotted lines have been drawn at the maximum density and initial blob positions, respectively. (An animation of this figure is available online. The animation runs for a total real-time of $660$ s. It shows the evolution of the variables presented in the figure and how the initial Gaussian profile of the blob, given by Eq. (\ref{eq:blob_profile}) develops into a V-shape as the plasma falls down.)}
    	\label{fig:maps}
    \end{figure*}

    Once the temporal evolution starts, the partially ionized blob begins to fall towards the lower boundary under the action of gravity. As described by \citet{Oliver2014ApJ...784...21O} the motion followed by the blob does not correspond to what is expected for a free-fall. This is due to the development of a pressure gradient that opposes the gravity force and slows down the plasma. The falling speed strongly depends on the blob to corona density ratio, with a larger density leading to a faster fall since a larger pressure gradient is required to balance the gravity force. Then, in agreement with the results from \citet{Martinez2020A&A...634A..36M}, the horizontal variation of the density leads to a deformation of the blob because its lighter regions fall slower than the denser ones. From its initial Gaussian profile, given by Eq. (\ref{eq:blob_profile}), the blob turns into an elongated V-shape as time advances. This deformation can be checked by looking at the density contours represented in Fig. \ref{fig:maps} (see also the animation accompanying this figure).
    
    The charged fluid is directly affected by the magnetic field while the neutral one is not, which in principle would lead to different dynamics for each fluid. Another reason for the different dynamics of the two fluids would be that their blob to corona density ratio and their vertical scale heights are different. However, in the simulation studied here both components of the plasma show an almost identical behavior due to the existence of the collisional coupling. Broadly speaking, both follow the evolution described in the previous paragraph, although important differences between them can be found. For instance, the color maps on panels \textit{a)} and \textit{b)} of Fig. \ref{fig:maps} show the $x$-component and the $z$-component of the drift velocity between the two fluids, respectively, at a given time of the simulation. In the central and denser part of the blob, the drift velocity is very small, of the order of centimeters per second, in agreement with the results from the 1D study of \citet{Oliver2016ApJ...818..128O}. This is a consequence of a very strong collisional coupling caused by a very large total density. Nonetheless, much larger drift velocities, of the order of kilometers per second are also present, which means that there are regions where the charged-neutral interaction is not as strong as in the core of the blob. We see that these large drift velocities appear mainly at the wings of the blob and at its external layers. The presence of such differences in the velocities of the two fluids is accompanied by heating of the plasma, as shown on panel \textit{c)} of Fig. \ref{fig:maps}, where we represent the frictional heating term given by Eq. (\ref{eq:qv}).
    
    \begin{figure*}
	    \centering
        \includegraphics[width=0.85\hsize]{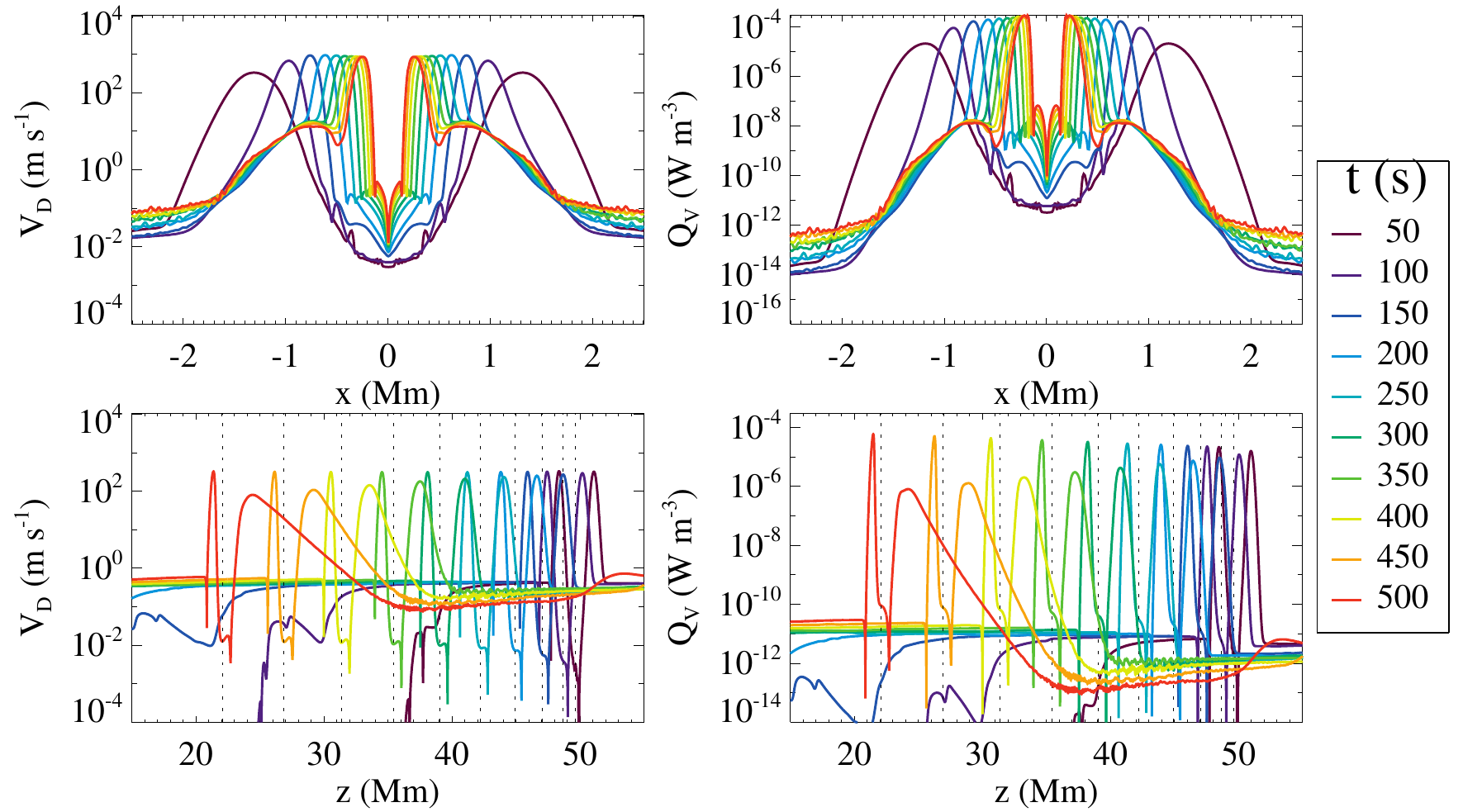}
	    \caption{Top and bottom panels show horizontal and vertical cuts, respectively, of the absolute value of the drift velocity (left) and the heating term (right). Each line color corresponds to a different time of the simulation. Vertical dotted lines in the bottom panels mark the position of the density peak at each time.}
	    \label{fig:cuts}
    \end{figure*}
    
    An interesting feature of the evolution of the falling blob is that the horizontal drift velocities are consistently larger than the vertical ones, as it can be checked by comparing the panels \textit{a)} and \textit{b)} of Fig. \ref{fig:maps}. This fact can be explained as follows. \citet{Martinez2020A&A...634A..36M} showed that, as a fully ionized plasma blob falls, the gas pressure tends to produce a horizontal expansion, which would lead to a large deformation of the falling condensation. However, in the presence of a relatively strong magnetic field there is a magnetic tension that almost perfectly balances the effect of the pressure. In the case of a partially ionized plasma, only the charged component is directly affected by this magnetic tension and a noticeable horizontal expansion of the neutral component is only prevented by the effect of the collisional coupling. Conversely, the influence of the magnetic field on the vertical motion is almost negligible, so there is not such a difference in the forces affecting each fluid. Therefore, in the horizontal direction a large drift velocity is required to play a similar role to the magnetic tension of the fully ionized scenario, while in the vertical direction smaller drift velocities appear.
    
    To get a more detailed picture of the evolution of the drift velocities and heating profiles, we show in Fig. \ref{fig:cuts} horizontal and vertical cuts of these variables at different times of the simulation. The horizontal cuts, shown in the top panels, are computed at the height where the maximum of density is located at each moment, represented in Fig. \ref{fig:maps} by the horizontal black dashed lines (the horizontal dotted lines in Fig. \ref{fig:maps} show the starting position of the blob). The position of the vertical cuts (bottom panels of Fig. \ref{fig:cuts}) is shown in Fig. \ref{fig:maps} as dotted black lines. We see in the horizontal cuts that the drift velocities and heating profiles (left and right panels, respectively) share their main features: very small values are found at the center ($x = 0$); then, there is a large increase in magnitude as we move towards the wings of the falling blob; and, finally, their magnitudes decrease again in the plasma of the external corona. As a consequence of the blob deformation during its fall, these horizontal cuts display two maxima that become less separated with time. The vertical cuts show that the larger drift velocities and heating appear at the front and the back of the blob, while they are negligible at its center. Figure \ref{fig:cuts} also reveals that the magnitudes of the drift velocities and the frictional heating increase with time. This behavior is discussed below.
    
    \begin{figure}
    	\centering
	    \includegraphics[width=\hsize]{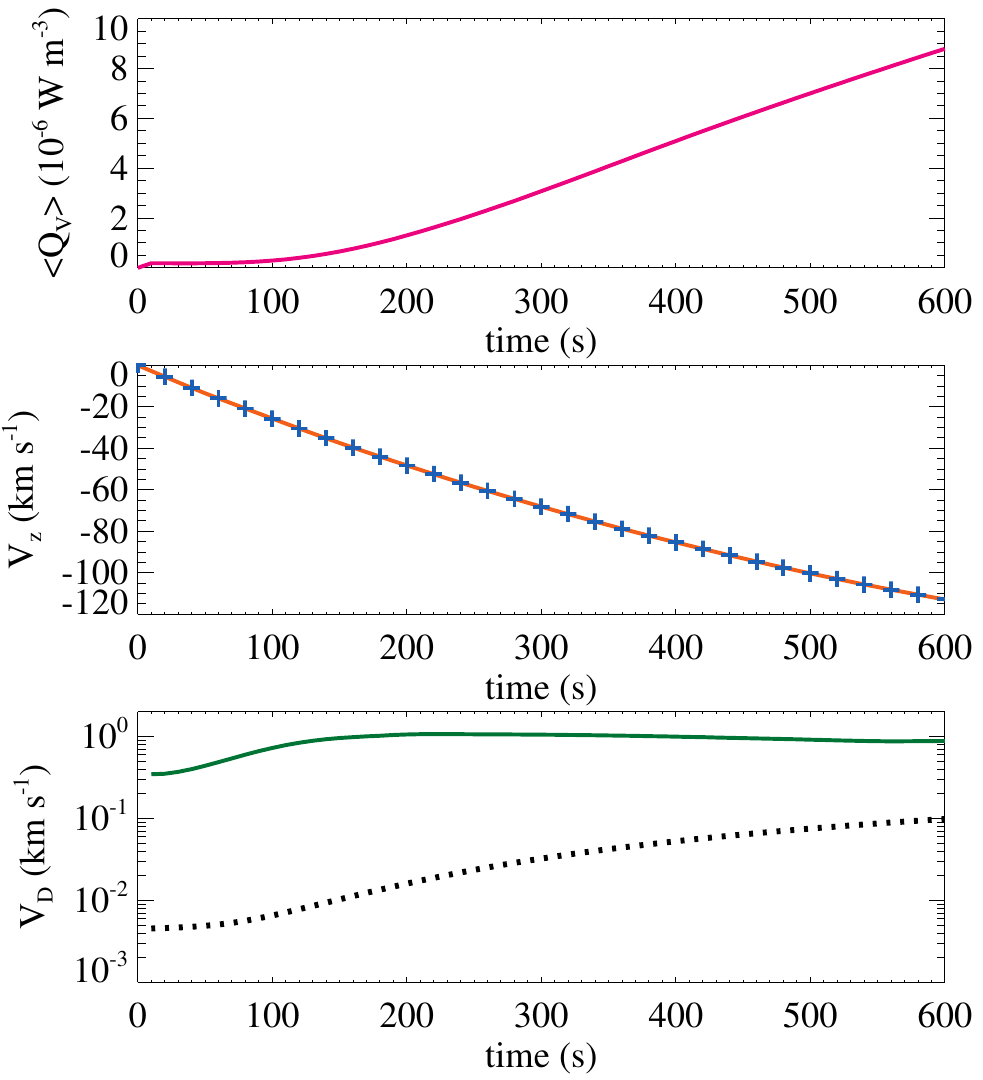}
	    \caption{Top panel shows the spatially averaged value of the heating term, $\langle Q_{V} \rangle$, as a function of time. The solid orange line and the blue crosses in the middle panel show the falling speed of the density peak of the charged and the neutral fluid, respectively, as a function of time. The bottom panel shows the maximum of the drift velocity (green solid lines) and the spatially averaged drift velocity (black dotted line) as a function of time.}
	    \label{fig:qmean}
    \end{figure}
    
    The increase of the frictional heating with time is shown in the top panel of Fig. \ref{fig:qmean}, where we represent the spatial average of this quantity as a function of time (the spatial averages are performed over the computational domain given by $-2.5 \ \rm{Mm} < x < 2.5 \ \rm{Mm}$ and $0 < z < 60 \ \rm{Mm}$, that is, the domain represented in Fig. \ref{fig:maps}). The main reason for this increase is that in this simulation the blob is still gaining speed as it falls. The pressure gradient has not been able to fully balance the gravity force yet and the blob has not reached the constant speed phase described in \citet{Oliver2014ApJ...784...21O} and \citet{Martinez2020A&A...634A..36M}. This behavior of the falling speed is represented in the middle panel of Fig. \ref{fig:qmean}. In addition, in the bottom panel of Fig. \ref{fig:qmean} we compare the temporal evolution of the maximum value of the drift velocity (green solid line) and the spatially averaged drift velocity, $\langle V_{\rm{D}} \rangle$ (dotted line). On the one hand, we see that the maximum drift velocity is of the order of $1 \ \rm{km \ s^{-1}}$, which is approximately $1 \%$ of the falling speed, while the spatially averaged drift velocity at the end of the simulation is of the order of $100 \ \rm{m \ s^{-1}}$. On the other hand, we see that the maximum of the drift velocity saturates around the time $t \sim 200 \ \rm{s}$, while the spatially-averaged values keeps increasing with time. This increase is caused by a larger fraction of the domain reaching the saturation value of drift velocity as time advances, and leads to the build-up of $\langle Q_{V} \rangle$.
    
    After analyzing the evolution of the drift velocities and the frictional heating, it is also interesting to investigate the variations of temperature related to the charged-neutral interaction. As shown by \citet{Martinez2021A&A...650A.123M}, the frictional heating produces an increase of temperature of each fluid given by the following expressions:
    \begin{equation} \label{eq:deltats}
        \delta T_{\rm{c,coll}} = \frac{\left(\gamma - 1 \right)}{n_{\rm{c}} k_{\rm{B}}}Q_{V} \ \ \text{and} \ \ \delta T_{\rm{n,coll}} = \frac{\left(\gamma - 1 \right)}{n_{\rm{n}} k_{\rm{B}}}Q_{V},
    \end{equation}
    where $\gamma = 5/3$ is the adiabatic constant. These expressions give a direct proportionality between the friction heating and the temperature increase of both species.
    
\begin{figure*}
	    \centering
	    \includegraphics[width=0.85\hsize]{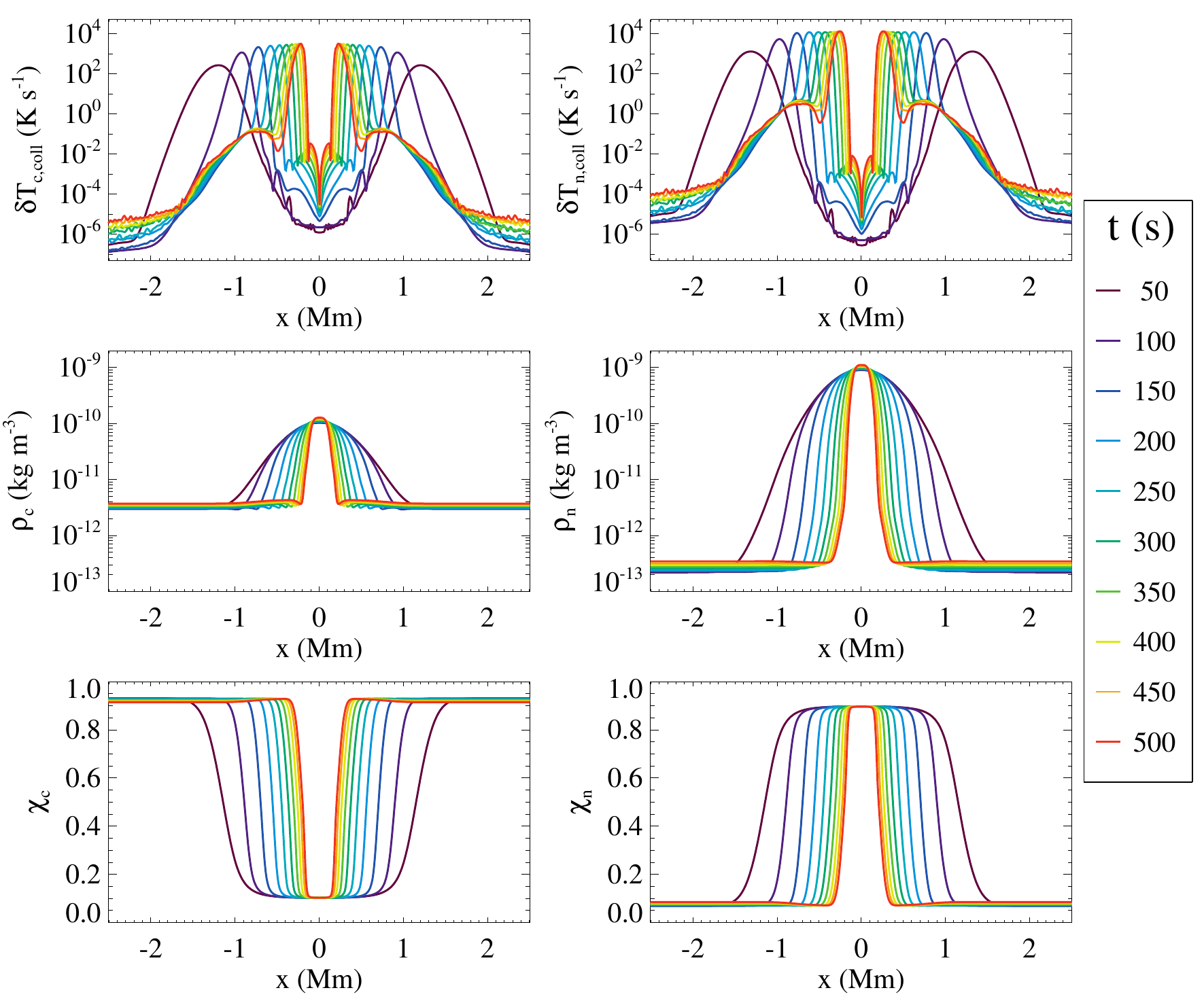}
	    \caption{Horizontal cuts of the temperature variations caused by charged-neutral collision (top panels), densities (middle panels) and relative abundances, $\chi_{\rm{c}}$ and $\chi_{\rm{n}}$ (bottom panels). Left panels show the results for the charged fluid and right panels show the results for the neutral fluid. Each line color corresponds to a different time of the simulation.}
	    \label{fig:cuts_temps}
    \end{figure*}

    We show in panels \textit{d)} and \textit{e)} of Fig. \ref{fig:maps} color maps of the variables $\delta T_{\rm{c,coll}}$ and $\delta T_{\rm{n,coll}}$, respectively, together with neutral density contours. As expected, the largest variations of temperature are located at the areas with the largest values of $Q_{V}$. To understand the temperature increase of both species, we compare the profiles of the variations of temperature with those of the densities and the relative abundances ($\chi_{\rm{c}}$ and $\chi_{\rm{n}}$). In the same way as in Fig. \ref{fig:cuts} for the cases of the drift velocities and the heating term, we show in Fig. \ref{fig:cuts_temps} horizontal cuts of the temperature variations (top panels), of the densities (middle panels), and of the relative abundances (bottom panels) at selected times of the simulation. Results for the charged and the neutral fluids are shown in the left and right panels, respectively. We see that the profiles of the temperature variations are similar to those of the frictional heating shown in Fig. \ref{fig:cuts}: the larger values appear at the wings of the blob while the variations of temperature at the core of the blob are negligible. In addition, at every step of the simulation the maximum values of $\delta T_{\rm{n,coll}}$ are larger than those of $\delta T_{\rm{c,coll}}$, in agreement with the results presented in Fig. \ref{fig:maps}. Then, the comparison with the density and abundance profiles shows that the regions of larger values of $\delta T_{\rm{c,coll}}$ and $\delta T_{\rm{n,coll}}$ correspond to regions with larger values of the ionization degree, that is, where the density of neutrals is much smaller than the density of charged particles ($\chi_{\rm{n}} \ll \chi_{\rm{c}}$). Finally, Fig. \ref{fig:cuts_temps} clearly shows that the most important increases of temperature due to the collisional interaction occur at the transition layer between the blob and the coronal plasma.
    
    Finally, to complement this analysis of the heating and temperature variations of the plasma, we refer the reader to Appendix \ref{sec:appendix}, where we present a comparison between the frictional heating due to collisions and the adiabatic heating caused by the compression of the plasma as the blob falls through the corona.
    
\section{Discussion and conclusions}
    In this paper we have used a two-fluid model to simulate the fall of a partially ionized coronal rain blob. Our results show the existence of several regimes of interaction between the charged and neutral species contained in the plasma. On the one hand, due to the large densities at the core of the blob, there is a very strong collisional coupling and, therefore, the dynamics of this region is well described by single-fluid model \citep{Oliver2016ApJ...818..128O,Martinez2020A&A...634A..36M}. On the other hand, as we move towards the external layers of the cold condensation, the density decreases and the coupling due to collisions becomes weaker. This is specially evident in the thin region that connects the weakly ionized blob with the fully ionized coronal plasma. The simulation analyzed here shows that, as a consequence of this weak coupling, large drift velocities between the charged and the neutral species appear, accompanied by heating of the plasma due to the collisional friction. This implies that there is no longer a single-fluid dynamics and that the use of a multi-fluid description is required. To the best of our knowledge, this is first time that the existence of large ion-neutral velocity drifts in coronal rain events is reported, either in a numerical or in an observational work.
    
    The maximum drift velocities found in our simulation are of the order of $1 \ \rm{km \ s^{-1}}$. 
    Similar values have been reported in other studies of partially ionized plasmas in different scenarios of the solar atmosphere, such as numerical simulations of the RTI \citep{Popescu2021A&A...650A.181P} or observations of solar prominences \citep{Khomenko2016ApJ...823..132K,Anan2017A&A...601A.103A,Stellmacher2017SoPh..292...83S,Wiehr2019ApJ...873..125W,Wiehr2021ApJ...920...47W,Gonzalez-Manrique2022,Zapior2022ApJ...934...16Z}. Therefore, the combination of numerical simulations and observations offers a coherent picture of the evolution of the partially ionized plasmas in coronal structures such as prominences or coronal rain blobs: while the various components of the plasma approximately behave as a single fluid in the central and denser regions of these structures, important differences in the motions of the charged and the neutral species appear at the edges. Coronal rain drops are expected to be essentially optically thin, which would facilitate measuring velocities of different species. An observational confirmation of the drift velocities predicted by the present numerical study could be obtained through a dedicated campaign on a large-aperture telescope. It should be taken into account that, according to the results presented in Fig. \ref{fig:maps}, it would be easier to detect the drifts in the horizontal motions of the plasma than along the direction of the fall of the blobs.
    
    In our simulation we have found that the charged-neutral interaction causes increases of temperature of up to $10^{3} - 10^{4} \ \rm{K \ s^{-1}}$. In principle, these values seem extremely high but we remind that they occur at the blob to corona transition regions, where the initial temperature of the plasma is already of the order of $10^{5}$ to $10^{6} \ \rm{K}$. Moreover, the magnitude of the blob temperature and its variation might not be accurate for a realistic scenario, since in our model we have not followed the full blob formation cycle \citep{Mikic2013ApJ...773...94M,Kohutova2020A&A...639A..20K,Li2022ApJ...926..216L} driven by the thermal instability \citep{Parker1953ApJ...117..431P,Field1965ApJ...142..531F}. For this research, we have resorted to the ideal equation of state and we have neglected non-adiabatic processes such as thermal conduction, radiative losses or ionization and recombination. Furthermore, we have started the simulation with an already formed blob which is included as a density perturbation superimposed to the background atmosphere. In this way, the initial temperature of the blob is directly related to the density ratio between the perturbation and the equilibrium. 
    
    Thus, our findings regarding the heating process should be taken as a qualitative description instead of a quantitative one. Nevertheless, from this qualitative point of view we may argue that the frictional heating obtained in the simulation would have associated observational signatures. It would produce an enhanced emission coming from the blob to corona transition region when observed in spectral lines associated to hot plasma, which might explain the brightenings detected in coronal rain observations \citep[see, e.g.,][]{Antolin2020PPCF...62a4016A}. This statement could be checked by performing forward modelling \citep[see, e.g.,][]{Antolin2022ApJ...926L..29A,Antolin2022FrASS...920116A} and comparing the results from simulations of a partially ionized plasma with those for the fully ionized case. However, this task is out of the scope of the present paper and is left for a follow-up research.  Finally, to get a more accurate description of this phenomenon it is important to include in future works the non-adiabatic effects that have not been considered here. For instance, the processes of ionization and recombination can enhance the density of the charged fluid in a thin layer around the cold blob, due to the ionization of the neutrals at the interface between the blob and the corona \citep{Popescu2021A&A...650A.181P}. As a consequence, an enhanced emission could be produced at this layer in a similar way as it has been detected in prominence to corona transition regions \citep{Berger2017ApJ...850...60B}. In addition, as shown by \citet{Zhang2021ApJ...911..119Z}, ionization and recombination generate larger decoupling between charges and neutrals, which would result in larger drift velocities and collisional heating.
    
\acknowledgements
    This work was supported by the European Research Council through the Consolidator Grant ERC-2017-CoG-771310-PI2FA. This publication is part of the R+D+i project PID2020-112791GB-I00, financed by MCIN/AEI/10.13039/501100011033. D.M. also acknowledges support from the Spanish Ministry of Science and Innovation through the grant CEX2019-000920-S of the Severo Ochoa Program. We thankfully acknowledge the technical expertise and assistance provided by the Spanish Supercomputing Network (Red Española de Supercomputacion), as well as the computer resources used: the LaPalma Supercomputer, located at the Instituto de Astrofisica de Canarias. We also would like to thank the anonymous referee for providing very useful comments.
	 
\bibliographystyle{aasjournal}
\bibliography{cr_bib}

\appendix
\section{Comparison of frictional and adiabatic heating} \label{sec:appendix}
    In the main text of this work we have only paid attention to the non-ideal process of plasma heating due to collisions between the charged and the neutral particles. However, our model also contains an ideal mechanism that plays a role in the variations of temperature of the two fluids: the adiabatic heating (and cooling) related to compressions (and expansions) of the plasma. In the present section we study how the ideal and non-ideal heating processes compare to each other.
    
    From \citet{Martinez2021A&A...650A.123M}, the equations that describe the temporal evolution of the internal energy of the charged and the neutral fluids, $e_{\rm{c}}$ and $e_{\rm{n}}$, respectively, are given (in the absence of Joule heating) by
    \begin{equation} \label{eq:ienec}
        \frac{\partial e_{\rm{c}}}{\partial t} + \nabla \cdot \left(e_{\rm{c}} \bm{V}_{\rm{c}} \right) + P_{\rm{c}} \nabla \cdot \bm{V}_{\rm{c}} = Q_{T} + Q_{V}
    \end{equation}
    and
    \begin{equation} \label{eq:ienen}
        \frac{\partial e_{\rm{n}}}{\partial t} + \nabla \cdot \left(e_{\rm{n}} \bm{V}_{\rm{n}} \right) + P_{\rm{n}} \nabla \cdot \bm{V}_{\rm{n}} = -Q_{T} + Q_{V},
    \end{equation}
    where $Q_{T}$ is the term of thermal exchange due to collisions, which depends on the difference of the temperatures of the two fluids and has the main effect of keeping the two components of the plasma at the same temperature. Neglecting the effect of this term, Eqs. (\ref{eq:ienec}) and (\ref{eq:ienen}) can be written as
    \begin{equation} \label{eq:ienec_2}
        \frac{\partial e_{\rm{c}}}{\partial t} + \nabla \cdot \left(e_{\rm{c}} \bm{V}_{\rm{c}} \right) = Q_{\rm{ad},c} + Q_{V}
    \end{equation}
    and
    \begin{equation} \label{eq:ienen_2}
        \frac{\partial e_{\rm{n}}}{\partial t} + \nabla \cdot \left(e_{\rm{n}} \bm{V}_{\rm{n}} \right) = Q_{ad,n} + Q_{V},
    \end{equation}
    where
    \begin{equation} \label{eq:adiabatics}
        Q_{\rm{ad},c} \equiv -P_{\rm{c}} \nabla \cdot \bm{V}_{\rm{c}} \quad \text{and} \quad Q_{\rm{ad},n} \equiv -P_{\rm{n}} \nabla \cdot \bm{V}_{\rm{n}}
    \end{equation}
    are the variations of the internal energy of each fluid related to the expansions and compressions of the plasma, that is, the adiabatic cooling and adiabatic heating, respectively.
    
     \begin{figure*}
    	\centering
    	\includegraphics[width=\hsize]{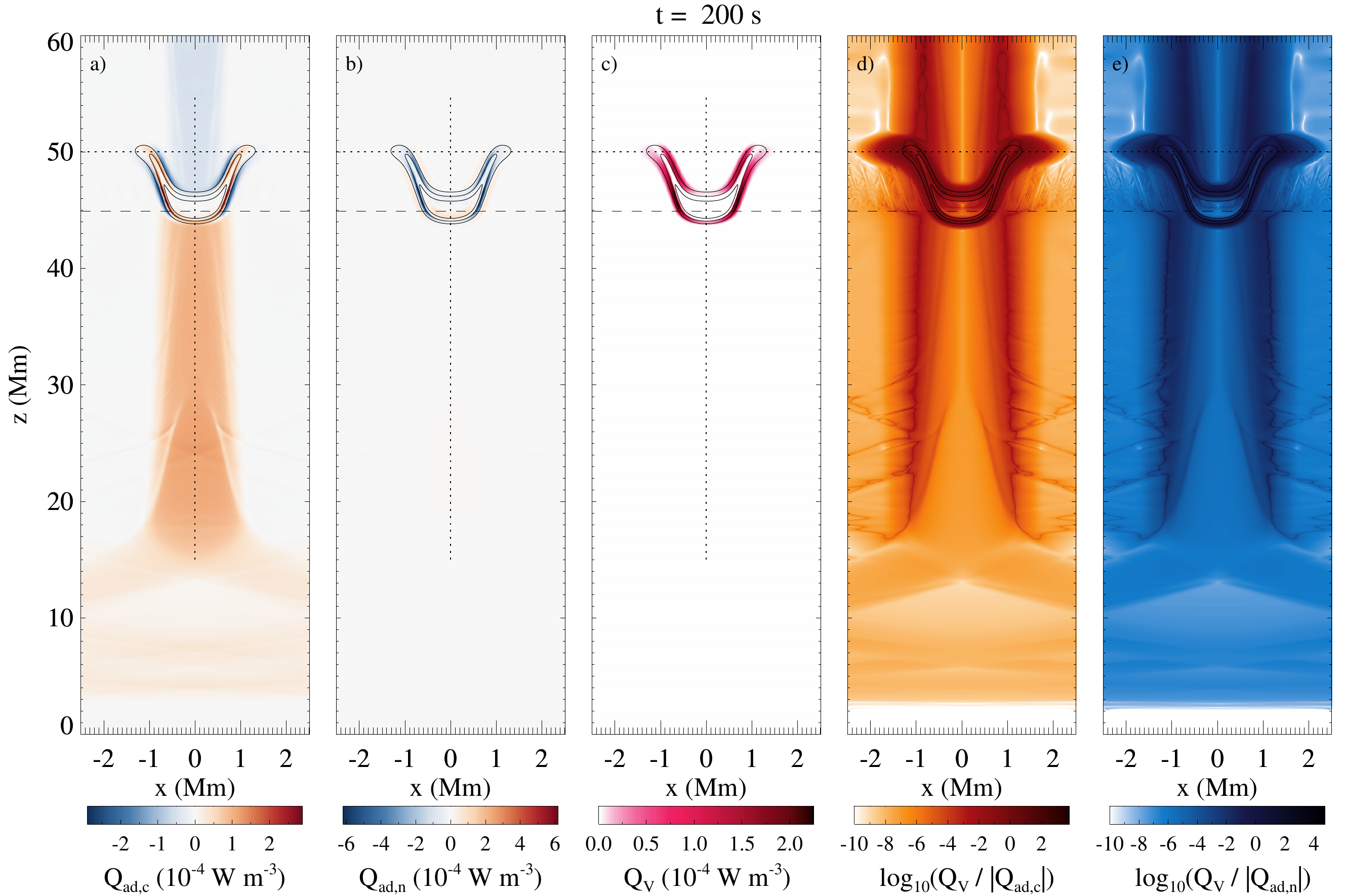}
    	\caption{Comparison between adiabatic and frictional heatings. Panels \textit{a)} and \textit{b)} represent the adiabatic variation of internal energy of the charged and the neutral fluids, $Q_{\rm{ad},c}$ and $Q_{\rm{ad},n}$, respectively. Panel \textit{c)} shows the frictional heating due to collisions. Panels \textit{d)} and \textit{e)} represent the ratios between the frictional heating and the absolute values of $Q_{\rm{ad},c}$ and $Q_{\rm{ad},n}$, respectively.}
    	\label{fig:heatings}
    \end{figure*}
    
    \begin{figure}
    	\centering
	    \includegraphics[width=0.5\hsize]{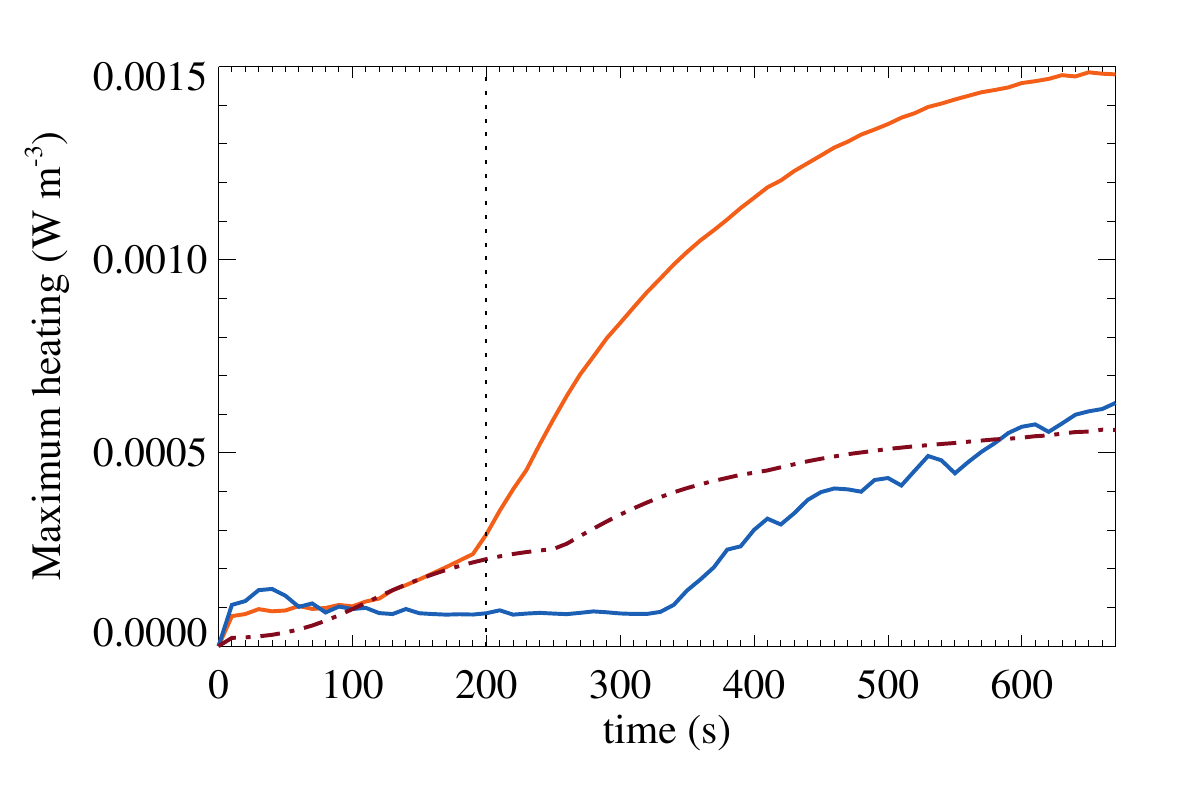}
	    \caption{Maximum values of the heating terms as functions of time. Orange and blue solid lines correspond to the maximum values of the adiabatic heating of the charged and the neutral fluid, respectively. The dotted-dashed line represent the results for the frictional heating due to charged-neutral collisions. The vertical dotted line shows the time of the simulation represented in Fig. \ref{fig:heatings}.}
	    \label{fig:maxheatings}
    \end{figure}
    
    We present in Fig. \ref{fig:heatings} the computations of the different kinds of heating terms for a given snapshot of the simulation analyzed in the main body of the paper. Panel \textit{a)} shows that, as the blob falls, the charged fluid below it increases its temperature because there is a compression caused by an increase of the pressure \citep[see][]{Martinez2020A&A...634A..36M}. The opposite situation occurs above the falling blob: the pressure reduces and, consequently, the plasma cools down. The same behavior takes place in the neutral fluid, although it cannot be clearly seen in panel \textit{b)} of Fig. \ref{fig:heatings} because of the large difference in the pressure of neutrals between the interior of the blob and the corona. Nevertheless, panels \textit{a)} and \textit{b)} show that there large variations of temperature of the two fluids in the borders of the blob. The comparison with the frictional heating shown in panel \textit{c)} reveals that the adiabatic and the frictional heatings are of the same order of magnitude around the blob but, more important, they do not take place at the same locations. Thus, there are regions of the domain where the only contribution to plasma heating comes from the charged-neutral collisions. This fact can also be checked by looking at the ratios $\log_{10} \left(Q_{V} / |Q_{\rm{ad,c}}| \right)$ and $\log_{10} \left(Q_{V} / |Q_{\rm{ad,n}}| \right)$ shown in panels \textit{d)} and \textit{e)}, where the darker colors correspond to areas in which the frictional heating dominates over the adiabatic mechanism (we note here that the color maps of these two panels are saturated in the lower limit of ratios, that is, for values below $-10$).
    
    Finally, Fig. \ref{fig:maxheatings} represents the maximum values of the heating terms as functions of time. To compute these values, we have considered the domain represented in Fig. \ref{fig:heatings}. We see that in general the adiabatic heating of the charged fluid is larger than the one for the neutral fluid and the frictional heating. However, the three terms are of the same order. In addition, we know from Fig. \ref{fig:heatings} that the largest values of the ratios $Q_{V} / |Q_{\rm{ad,c}}|$ and $Q_{V} / |Q_{\rm{ad,n}}|$ are located in the corona, not in the blob. Therefore, for the most part of the simulation the peak of the adiabatic heating of the charged fluid takes place far from the blob.
    
    In summary, the conclusion from this comparison is that the frictional heating is in general not negligible with respect to the adiabatic heating and that in certain regions of the domain (such as the borders of the falling blob) it is the main source of plasma heating according to our physical model.

\end{document}